# Automating Defense Against Adversarial Attacks: Discovery of Vulnerabilities and Application of Multi-INT Imagery to Protect Deployed Models

Josh Kalin*[a,b], David Noever[b], Matt Ciolino[b], Dominick Hambrick[b], Gerry Dozier[a]
[a]Department of Computer Science and Software Engineering, Auburn University, Auburn, AL, USA; [b]PeopleTec, Huntsville, AL, USA

**ABSTRACT**

Image classification is a common step in image recognition for machine learning in overhead applications. When applying popular model architectures like MobileNetV2, known vulnerabilities expose the model to counter-attacks, either mislabeling a known class or altering box location. This work proposes an automated approach to defend these models. We evaluate the use of multi-spectral image arrays and ensemble learners to combat adversarial attacks. The original contribution demonstrates the attack, proposes a remedy, and automates some key outcomes for protecting the model's predictions against adversaries. In rough analogy to defending cyber-networks, we combine techniques from both offensive ("red team") and defensive ("blue team") approaches, thus generating a hybrid protective outcome ("green team"). For machine learning, we demonstrate these methods with 3-color channels plus infrared for vehicles. The outcome uncovers vulnerabilities and corrects them with supplemental data inputs commonly found in overhead cases particularly.

## 1. INTRODUCTION

Image classifiers use deep neural networks to derive features and learn class labels for each dataset. In recent literature, adversarial patch camouflage has been used to evade detection from machine learning-based detectors. These works use knowledge of the underlying model architecture to create specialized patterns and masks that perturb or erase the necessary feature patterns for the classifier. This work proposes the use of Multi-INT imagery to overcome adversarial attack patterns. For instance, a vehicle draped with an adversarial patch can provide a heat signature in a repeatable, recognizable pattern. A classifier that uses both visible and infrared-based channels can distinguish the vehicle in experiments.

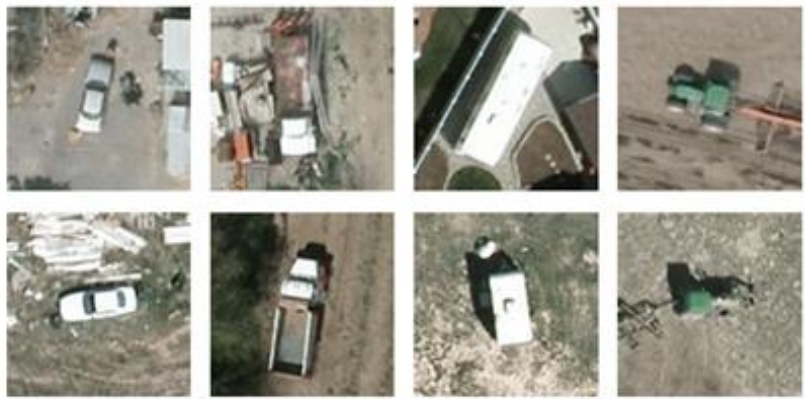

Figure 1. Vehicle Detection in Aerial Imagery (VEDAI) sample data showing different vehicles in different orientations from an overhead visible sensor.

**1.1** The VEDAI or Vehicle Detection in Aerial Imagery dataset is a benchmarking dataset with multiple channels looking at the same locations. The dataset includes nine vehicle classes that can be challenging to distinguish for computer vision applications. A critical aspect of this data is the rectified Infrared images included inside of this dataset. The original paper by Razakarivony and Jurie highlights the use of common machine learning applications to identify the classes inside of the dataset [1]. In this paper, the goal is to extend previous work by protecting a common image classification machine learning model from adversarial attacks. Three primary channels are included in the dataset: visible, infrared, and gray. For the experiments included in this paper, the Visible channel is also split into its principal components of red, green, and blue channels to explore the color dependence of the modeling.

**1.2** MobileNetV2, Inverted Residuals, and Linear Bottlenecks is a recent architecture improvement to the MobileNet design [2]. The newest paper offers greater accuracy vs. speed and accuracy vs. size comparisons to other state-of-the-art models. Tensorflow and other machine learning libraries offer methods for training and shrinking the models down for deployment on Single Board Computers / ASICs. This model is an ideal benchmark for exploration in the overhead imagery world given its superior accuracy, size, and compute requirements.

**1.3** Adversarial attacks are attacks on machine learning models that can manipulate or avoid detection. These algorithms are typically divided into two types: white-box and black-box attacks. With white-box attacks, the attacker knows the underlying training dataset and model architecture. Conversely, with black-box attacks, the attacker does not know any information about the underlying machine learning models. Universal black box attacks are rare, especially those that can work across multiple model types in a particular machine learning field. For the experiments conducted here, the focus is on using White box attacks with a library called FoolBox. FoolBox uses the base machine learning model and dataset to create tailored attacks to that model [3].

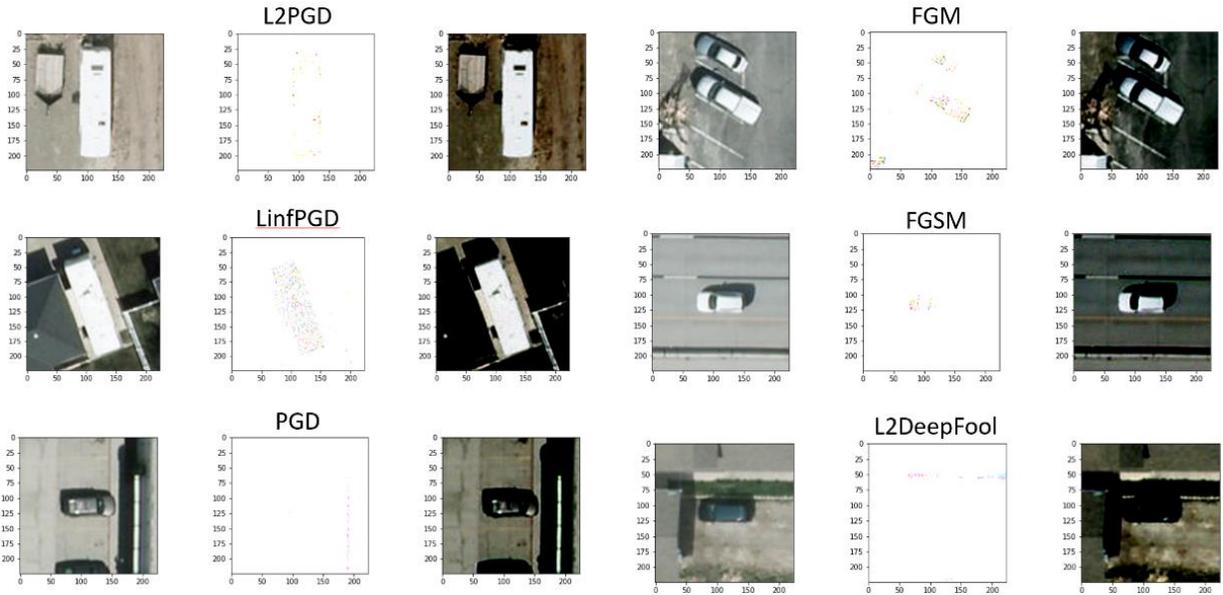

Figure 2. An example of the mask generated by Foolbox for each of the attack types. For each attack, on the left is the original image, the middle is the attack mask, and the right is the combination of the original and masks.

The FoolBox library has access to state-of-the-art adversarial attacks which including six types used in the experiments: Fast Gradient Sign Method, Fast Gradient Method, Projected Gradient Descent (PGD/Linf/L2), and L2DeepFool.

**1.4** FGSM: Fast Gradient Sign Method

Computes the gradient $g(x_0) = \nabla_x L(x_0, l_0)$ once and then seeks the minimum step size $\varepsilon$ such that $x_0 + \varepsilon sign(g(x_0))$ is adversarial. (Goodfellow et al., 2014 [4]). This method is fast and common.

**1.5 FGM: Fast Gradient Method**

Extends the FGSM to other norms [from infinity norm] and is therefore called the Fast Gradient Method. (adversarial-robustness-toolbox [5]). This method is also fast and common.

**1.6 PGD: Projected Gradient Descent**

Projected Gradient Descent (PGD) [6] is also an iterative extension of FGSM and very similar to the Basic Iterative Method [7] (BIM). The main difference with BIM resides in the fact that PGD projects the attack result back on the $\varepsilon$-norm ball around the original input at each iteration of the attack.

**1.7 PGD (LinfPGD, L2 PGD): Linf/L2 Projected Gradient Descent**

Linf Projected Gradient Descent is a PGD attack with $order = Linf$. L2 Projected Gradient Descent is a PGD attack with $order = L2$.

**1.8 L2DeepFool**

In each iteration, DeepFool (Moosavi-Dezfooli et al., 2015 [8]) computes for each class $l \neq l_0$ the minimum distance $d(l, l_0)$ that it takes to reach the class boundary by approximating the model classifier with a linear classifier. It then makes a corresponding step in the direction of the class with the smallest distance. (Rauber et al. 2017 [3])

**1.9 Securing Deployed Models Against Adversarial Attacks**

Recent papers focus on securing models during the development process. Green team machine learning creates a process called "Build, Attack, Defend" to evaluate the machine learning models during the development process and begin protecting against red team style attacks on the models [9]. Attacks are increasingly sophisticated with their ability to detect the underlying model architecture and therefore, exploit vulnerabilities in these models. Decision-making applications, like Automated (or Aided) Target Recognition (ATR), requires the ability of developers to reduce adversarial risk in their systems. Using a "Build, Attack, Defend" development process can lead to more secure models before deployment into production.

## 2. METHODS

**2.1** Automating protections for machine learning models involves a few steps. First, an automated evaluation pipeline should be created to evaluate the performance of the models against adversarial attacks. The methods section will cover the model types and adversarial attacks, the evaluation framework, and simple combinations of Multi-INT imagery to protect the models. "Build, Attack, Defend" systems for adversarial defenses are a primary inspiration for this work. The experiment creates baseline models, evaluates those models against a suite of adversarial attacks, and then proposes changes to defend the models from incoming attacks. A tiered protection system is introduced to segment the protections needed for the models.

**2.2** To understand model heritage, the experiments must observe the underlying training data and the model architecture. There are two primary questions:
- Is the deployed model trained on custom or novel data?
- Is the model trained from common datasets found in open source?

The concern with "download and deploy" strategies are that an attacker already has access to the architecture and the weights of the model – these types of attacks are called white-box attacks. It is much easier to construct attacks when the adversary has access to this information. In contrast, without access to the model or data (black-box attacks), the attacker will construct generic attacks and probe the model. With image classification, it is much more difficult to get feedback on what objects are being classified and tracked when there is no direct access to the model. In black box to white box, Ciolino et al. (2021) were able to detect the underlying learner architecture above 95% confidence [10]. Once a model type like MobileNetV2 is detected, a set of structured attacks can be evaluated on the models. The experiments included in this work will only focus on covering white-box attacks.

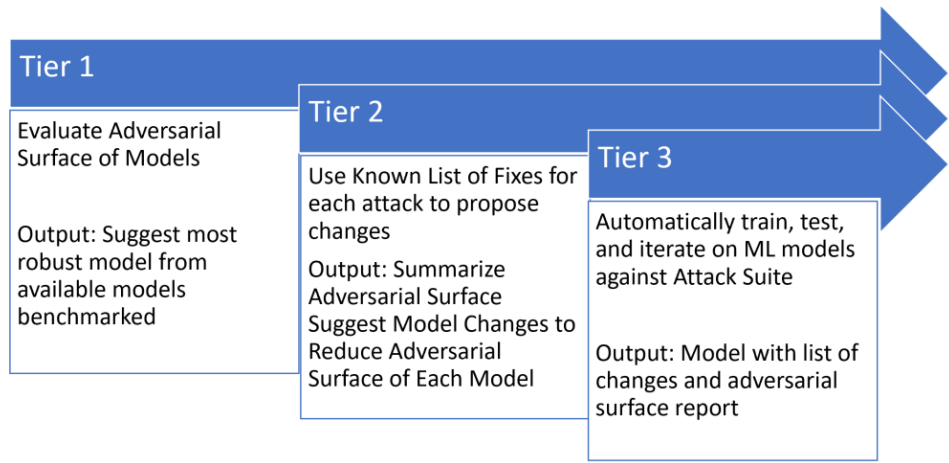

Figure 3: A tiered structure approach to adversarial defenses for machine learning models. Each tier represents additional work to protect the model from outside attackers.

**2.3** The evaluation framework contains a workflow seen in Figure 3 that provides an adversarial attack surface for a given model. There is a static set of modified images that are provided to the model and the detections are provided back for evaluation. Each set of images probes different adversarial susceptibility for the model. The product from this step is a report that details the susceptibility of the model to specific adversarial attacks – the adversarial surface. This is visualized with a Sankey diagram to highlight overall vulnerabilities for an analyst.

**2.4** After the adversarial surface is identified, the last tier is to retrain the model to reduce the efficacy of those attacks. There are three tiers currently in development for this automated exploration system. Each tier represents a way to secure the model. With every additional tier, there is a degree of complexity to protecting the model. Tier 1 protections involve evaluating a set of models and providing recommendations on which models are least susceptible to attack. Tier 2 takes the models and benchmarks them against a wide range of attacks to understand vulnerable areas and provides a report back on the adversarial surface of the current model. Finally, Tier 3 protections will automatically do architecture search and parameter tuning to reduce the overall efficacy of the test quite until user-defined thresholds are met. Recent work presented in adversarial camouflage demonstrates an evolving set of threats for the overhead machine learning [13]. This proposed tiered protection system provides a structured approach to vetting and fixing known issues.

**2.5** This work focuses on designing a system for aerial imagery that can provide Tier Level 1 results for Automated Protection of these machine learning systems. The underlying AI system will recommend which models provide the best robustness to adversarial attacks between all the methods and channels surveyed. The advantage of this exploration is that many overhead systems will have access to at least one visible channel and one additional imagery channel for detection. In Tier 2 experiments, this work demonstrates the adversarial surface of the models trained.

## 3. RESULTS

**3.1** The Tensorflow-Lite library offers simple MobileNetV2 training classes for easily building, benchmarking, and deploying classification models on low-end computing devices[14]. Using this library, each model is trained on an individual channel (VIS, Gray, IR, Red, Green, Blue) for 20 epochs with a batch size of 32. The images are 224 by 224, which is a typical training size for deep classification models.

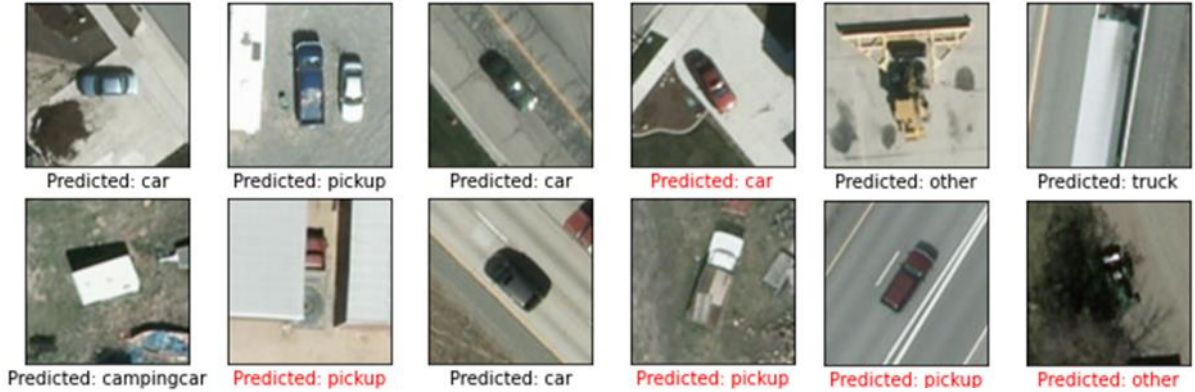

Figure 4. Randomly selected samples from the dataset with predictions from the Visible Model. The black predictions are correctly classified images and the red predictions are the incorrect predictions with the selected class.

MobileNetV2 is a general use case tool for deep classification tasks and the parameters can be tuned to provide the best performance. For this exploration, the experiments minimized the number of perturbations to keep a level playing field between the different models. Certain hyperparameters would provide advantageous against adversarial attacks. These ablations are left for future work on aerial imagery.

### 3.2 Training and Test Accuracies

For the MobileNetv2 architecture in Table 1, the training accuracy for the models hovers right around 90% for each of the channels and test accuracy is at the 80% mark. Delta accuracy for each model is not more than 5% offering similar performance between channels for the adversarial exploration in Section 3.2.

Table 1. Training and test accuracy for individual channels

| Channel | Training Accuracy | Test Accuracy |
| --- | --- | --- |
| Visible | 90.5% | 76.4% |
| Blue | 88.1% | 78.0% |
| Gray | 88.7% | 77.0% |
| Green | 90.2% | 79.8% |
| IR | 87.5% | 73.6% |
| Red | 88.7% | 78.9% |

### 3.3 Attack the models

The goal, in this section, is to understand the susceptibility of a model to attacks generated specifically for it or by an attacker who used a different channel. For instance, if an adversarial attacker used visible imagery as input to their attack generator, would the same attacks translate to IR? Is IR more robust against attacks generated by certain channels? This analysis explores the ability of an attack suite like Foolbox to create adversarial image generators that can attack both their matched channel and the other models from the same images.

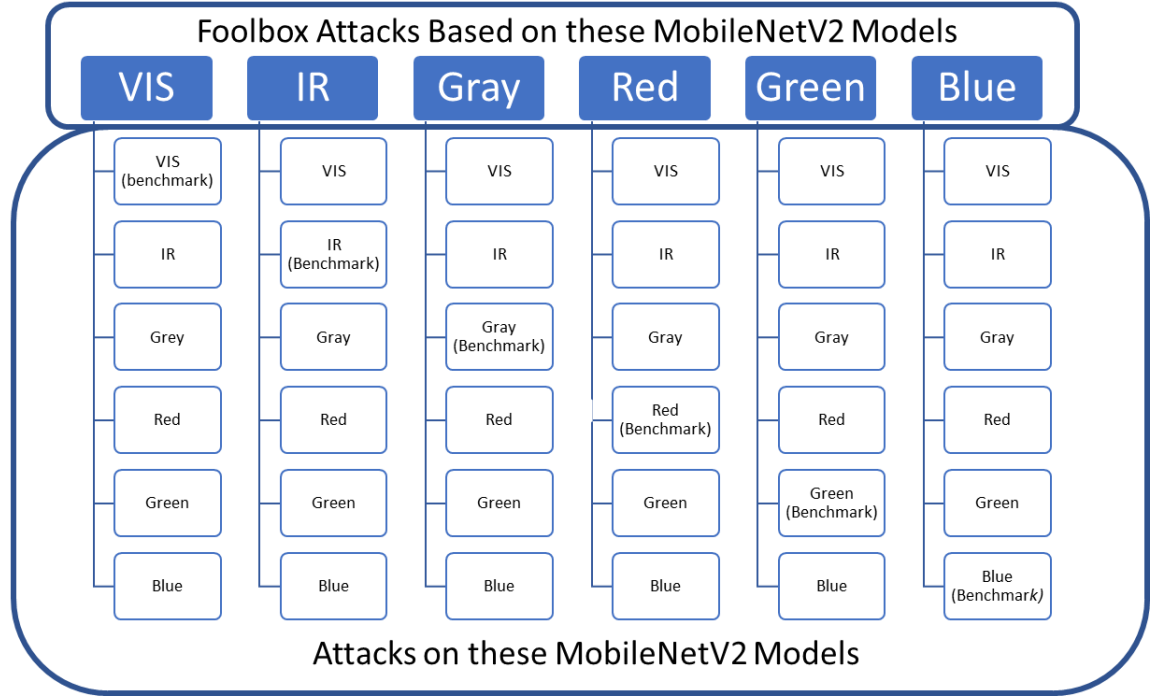

Figure 5. Experimental Design for Foolbox Attacks on VEDAI dataset. A Foolbox Attacker is trained on a single channel. The Foolbox model generates attacks for each channel and the adversarial efficacy is measured as a delta accuracy from original model accuracy to attacked model accuracy.

**3.4** Attacks Trained on a single channel and then used on all downstream models

The general process for evaluating the adversarial surface has the following steps:

1. Select a Benchmark Channel for the Attacks to be based on (White Box Attack)

    a. For example, let's select the Visible channel to train Foolbox.

2. Use the Adversarial Attack Generator to Attack the Benchmark Channel

    a. The Visible Adversarial Attack Generator generates poisoned images and feeds them into the Visible MobileNetV2 Model

    b. Measure the delta accuracy between the original inputs and poisoned inputs

    c. This is the benchmark data for Step 3 comparisons

3. Use the Adversarial Attack Generator to attack the other channels

    a. For each channel in [blue, green, red, gray, IR]

        i. The Visible Adversarial Attack Generator generates poisoned examples of the channel and feeds them to the channel model

        ii. Measure delta accuracy from original samples to the poisoned examples run through the selected model

4. Measure the Adversarial Surface of each model for each attack

This process provides a way to quantify the ability of each classification model to handle adversarial attacks from the benchmark and attacks that were trained on similar data.

### 3.5 Observations

The output of the adversarial surface evaluation has two primary visualizations – the overall surface and the per attack evaluations. Each one of these graphics shows the effectiveness of an attack. FGM and FGSM attacks can be mitigated through the appropriate selection of a model. For instance, a model trained on IR is resilient to attacks from FGM, and FGSM trained attackers on any of the other channels. Figure 6 shows the adversarial surface of the MobileNetV2 trained models against the Foolbox attackers.

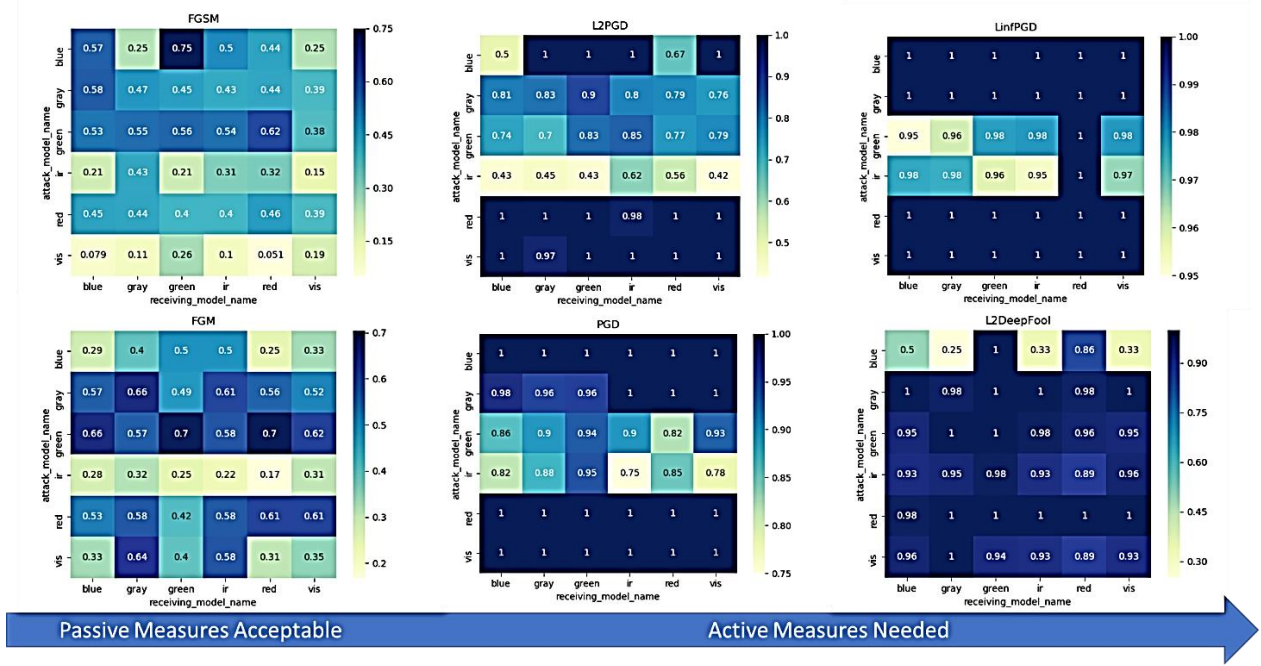

Figure 6. Adversarial Surface for MobileNetV2 Models trained on VEDAI dataset. Passive Measures included recommending a particular channel as mitigation. Active Measures require augmentation, architecture changes, and other changes to the modeling process to overcome the attacks.

### 3.6 Adversarial Transferability between channels

Transferability between the different imagery channels suggests that different colors learn separate features within a given architecture. For the visible images, their network will also learn individual features from each channel and use those features to distinguish between classes. For the single-channel training, the model will only have access to a single channel's features and will learn different weights. In the experiments, there are a few notable outcomes:

- Attacks trained with visible images and FGSM show the least reduction in model efficiency
- FGSM and FGM had the least successful attacks
- Four of the six attack types resulted in 100% detection in classification accuracy for most channels

The four attack methods that require active measures in the development process are left up to future work on this topic.

### 3.7 Recommendations from Adversarial Surface Data

The passive measure attackers can be mitigated through recommendations on available models evaluated. Because there are six available models, it is possible to use the adversarial surface to create a set of recommendations for which model should be used to protect against these types of adversarial examples. In this evaluation, FGSM and FGM are identified as attack sets that can mitigate by recommending an image classifier trained with the IR band.

**3.8** Future Work

Future work will focus on automating an end-to-end pipeline from dataset to model evaluations to a more robust model. Each of these individual pieces alone can protect a model. An automated system would optimize the model for maximum robustness to external attackers. There are trade-offs to a more robust model including complexity, accuracy, and explain-ability. A sample trade-off analysis can be included with the results.

## 4. CONCLUSIONS

This work demonstrates the ability to protect machine learning models for overhead imagery by simply using a structured approach to evaluate and reduce the adversarial surface of the machine learning model. The experiments demonstrate a reduction in the efficiency of adversarial attacks while maintaining original performance benchmarks. One critical issue to address from this work is how to reduce the computational burden of retraining the networks for every additional adversarial perturbation? In addition, further automation to protect models using the adversarial attack surface will be included in future works.